\documentclass{article}

\usepackage{amsmath}
\usepackage{graphicx}

\usepackage[a4paper, total={6in, 8in}]{geometry}
\usepackage{authblk}
\usepackage[colorlinks]{hyperref}
\usepackage[backend=bibtex]{biblatex}
\title{{Deep learning Based Correction Algorithms for 3D Medical Reconstruction in Computed Tomography and Macroscopic Imaging}} 

\author[1]{Tomasz Les}
\author[1,2]{Tomasz Markiewicz}
\author[3]{Malgorzata Lorent}
\author[2]{Miroslaw Dziekiewicz}
\author[1]{Krzysztof Siwek\thanks{Corresponding author: krzysztof.siwek@pw.edu.pl}}
\affil[1]{University of Technology, 00-661 Plac Politechniki 1, Warsaw, Poland}
\affil[2]{Military Institute of Medicine, 04-141, Szaserów 128, Warsaw, Poland}
\affil[3]{Institute of Tuberculosis and Lung Diseases, 01-138, Plocka 26, Warsaw, Poland}

\begin{document}

\maketitle

\begin{abstract}
{This paper introduces a hybrid two-stage registration framework for reconstructing three-dimensional (3D) kidney anatomy from macroscopic slices, using CT-derived models as the geometric reference standard. The approach addresses the data-scarcity and high-distortion challenges typical of macroscopic imaging, where fully learning-based registration (e.g., VoxelMorph) often fails to generalize due to limited training diversity and large nonrigid deformations that exceed the capture range of unconstrained convolutional filters. 
In the proposed pipeline, the Optimal Cross-section Matching (OCM) algorithm first performs constrained global alignment—translation, rotation, and uniform scaling—to establish anatomically consistent slice initialization. Next, a lightweight deep-learning refinement network, inspired by VoxelMorph, predicts residual local deformations between consecutive slices. The core novelty of this architecture lies in its hierarchical decomposition of the registration manifold: the OCM acts as a deterministic geometric anchor that neutralizes high-amplitude variance, thereby constraining the learning task to a low-dimensional residual manifold. This hybrid OCM + DL design integrates explicit geometric priors with the flexible learning capacity of neural networks, ensuring stable optimization and plausible deformation fields even with few training examples.
Experiments on an original dataset of 40 kidneys demonstrated that the OCM + DL method achieved the highest registration accuracy across all evaluated metrics: NCC = 0.91, SSIM = 0.81, Dice = 0.90, \mbox{IoU = 0.81}, \mbox{HD95 = 1.9 mm}, and volumetric agreement DC\(_{\text{Vol}}\) = 0.89. Compared to single-stage baselines, this represents an average improvement of approximately 17\% over DL-only and 14\% over OCM-only, validating the synergistic contribution of the proposed hybrid strategy over standalone iterative or data-driven methods. The pipeline maintains physical calibration via Hough-based grid detection and employs Bézier-based contour smoothing for robust meshing and volume estimation. 
Although validated on kidney data, the proposed framework generalizes to other soft-tissue organs reconstructed from optical or photographic cross-sections. By decoupling interpretable global optimization from data-efficient deep refinement, the method advances the precision, reproducibility, and anatomical realism of multimodal 3D reconstructions for surgical planning, morphological assessment, and medical education.}
\end{abstract}

\section{Introduction}
{In recent decades, advances in the field of medical imaging have significantly contributed to the development of diagnostics and treatment of kidney diseases. Technologies such as computed tomography (CT) and macroscopic imaging have become indispensable tools in identifying and analyzing kidney pathologies. However, despite the advanced capabilities of these methods, challenges related to geometric differences between 3D models obtained from various imaging sources exist. These differences can significantly impact the accuracy of diagnostics, surgical treatment planning, and medical education.}

{The aim of this study is to develop and evaluate geometric correction methods for 3D kidney models reconstructed from macroscopic imaging, using CT-based models as the reference standard.}

{In this work, computed tomography (CT) data is used as the anatomical reference standard, while the proposed correction methods are applied exclusively to 3D kidney models reconstructed from macroscopic imaging. The goal is to improve the accuracy and anatomical consistency of macroscopic models by aligning them with CT-based reconstructions.}

{In particular, 3D kidney models obtained via computed tomography offer high accuracy and detail thanks to the use of advanced image reconstruction algorithms. They allow for precise representation of the anatomical and pathological structure of organs. On the other hand, macroscopic models, created based on a series of photographic images of organ sections, can be less accurate due to potential tissue deformations, errors in matching sections, and physiological shrinkage of the extracted organs---for example, due to water loss. The combined use of translation, rotation, and scaling allows effective 3D alignment, compensating for their global effects and improving anatomical consistency. These differences can lead to errors in the interpretation of imaging data, directly affecting the quality of diagnostics and planning of surgical interventions.}

{In response to the geometric inconsistencies observed between both modalities, this article presents a developed method for the geometric correction of 3D kidney models obtained from macroscopic imaging. This method aims to minimize the differences between models obtained from different imaging techniques, through precise matching and geometric correction of macroscopic models.}

{By applying advanced image processing techniques and optimization algorithms, it is possible to obtain macroscopic models with significantly better accuracy and anatomical fidelity. Such an approach opens up new possibilities in the more accurate modeling of anatomical structures, which is crucial not only in diagnostics and surgery but also in scientific research and medical education.}

 {This work introduces a hybrid OCM + DL framework that decomposes registration into deterministic global alignment and learned residual refinement. This strategy ensures high accuracy and data efficiency on small clinical datasets ($N=40$), bridging the gap between macroscopic photography and CT standards while maintaining physical metric consistency.}

\section{Problem Statement}
The primary issue addressed in this article concerns the discrepancy in geometric accuracy between 3D kidney models obtained from computed tomography (CT) and those derived from macroscopic imaging. These differences can lead to inconsistencies in the interpretation of imaging data, which directly impact the effectiveness of diagnostics, surgical planning, and medical education in the field of kidney anatomy. This issue is particularly significant in the context of precisely modeling pathological changes, such as kidney tumors, where the accuracy of the model is crucial for treatment success \cite{Pennarossa2021, Wang2017, Soler2014, Hu2007, Geiger1993, abdullah2018development}.

The challenge lies in developing a method capable of effectively correcting these differences, enabling the creation of accurate and anatomically faithful 3D kidney models based on macroscopic data. Such a method must be able to address potential tissue deformations and errors in matching sections, which are typical for the process of creating macroscopic models. This solution must also be sufficiently flexible to be applied to modeling different types and sizes of kidneys, as well as varying degrees of pathology progression \cite{zhang20203d, sarmah2023survey}.

\begin{figure}[ht]

\includegraphics[width=0.8\linewidth]{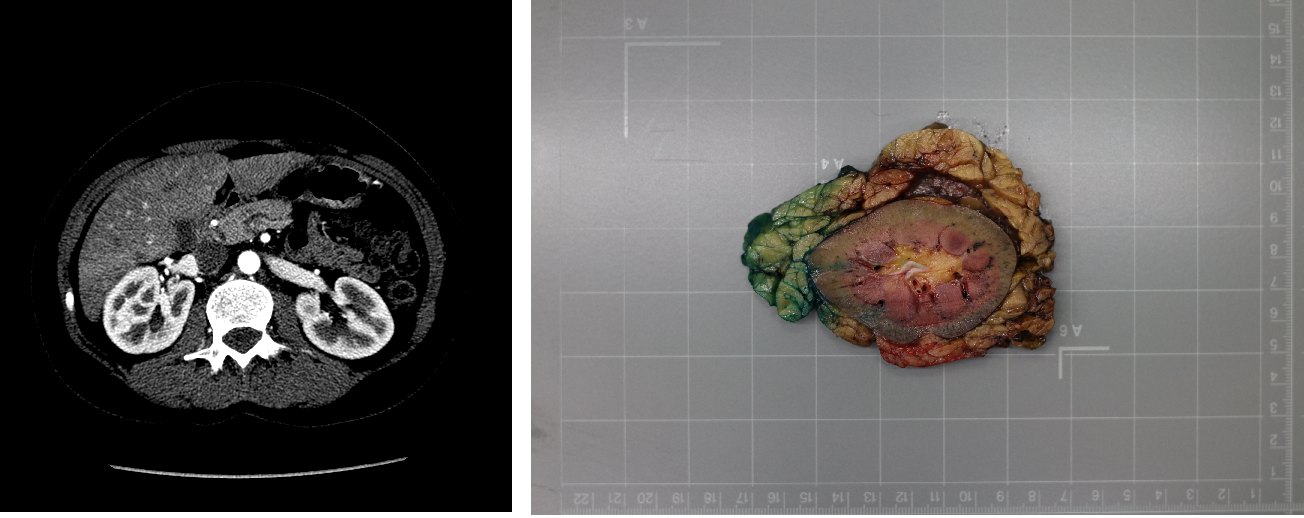}
\caption{{On the} 
 (\textbf{left}), {a CT}
{scan} 
 section of a kidney is shown, displaying the organ's detailed internal structure. On the (\textbf{right}), a macroscopic image of a kidney is presented, illustrating its external appearance and shape, providing information on the organ's surface condition.}
\label{fig:ctimacro}
\end{figure}

In the context of this challenge, we present a CT image of a kidney and a macroscopic scale image of a kidney in Figure \ref{fig:ctimacro}. Due to differences in scale, sample preparation methods, and the inherent nature of both imaging techniques, these sections can vary significantly.

These differences are not limited to visual aspects but also have a profound impact on the process of creating 3D models. In particular, when constructing 3D kidney models from both CT and macroscopic data, the differences can amount to up to 30\% in volume and in specific dimensions, such as length, width, and thickness. Recent comparative studies confirm that macroscopic observation and digital 3D reconstructions (e.g., CT or surface scans) often yield inconsistent measurements and interpretations due to imaging artifacts, handling differences \cite{abegg2021virtual, viero2021potential}. Such discrepancies pose a significant challenge to the accuracy of anatomical modeling, directly affecting the reliability of the models used in medicine, which underscores the importance of developing an effective method for correcting these differences.

Previous approaches to macroscopic imaging quantification have typically relied on manual or semi-manual scaling using visible reference markers such as rulers or calibration grids placed in the photographic background. These methods often involve estimating pixel-to-millimeter conversion factors by hand, which introduces variability and is sensitive to camera angle, lighting, and slice deformation. Despite their simplicity, such techniques remain widespread due to the lack of standardized, automated tools. Our approach enhances this process by introducing automated detection of scale features using the Hough Transform and by integrating slice-wise geometric alignment to improve consistency and accuracy across reconstructed 3D volumes.

Recent advancements in computational methods and deep learning algorithms have catalyzed a significant evolution in the realm of 3D reconstruction from computed tomography (CT) scans. Contrary to earlier limitations, contemporary research highlights the integration of sophisticated image processing techniques and machine learning models, marking a pivotal enhancement in the precision of anatomical modeling. There are numerous studies on automated CT organ detection systems, which exemplify the integration of advanced computational techniques to improve the accuracy and efficiency of medical diagnostics \cite{Hagen2021, Chun2022, Wang2018, Rister2020, Armato2003}. This trend is reinforced by comprehensive reviews summarizing the state of the art in 3D medical reconstructions, including both classical and deep learning-based methods \cite{li2024comprehensive}.

The development of CNN-based segmentation for cardiac micro-CT data, for instance, exemplifies the push towards utilizing deep learning for enhancing the granularity and accuracy of 3D reconstructions, underscoring the method's capability in high-throughput phenotyping applications \cite{badea2024whole}. Similarly, innovations in scaffolding methods for cartilage restoration further illustrate the multifaceted applications of 3D CT reconstructions in regenerative medicine, emphasizing its role in fabricating conducive chondrogenic environments~\cite{zhao2024hybrid}. 

A critical point in recent research is the correction of geometric artifacts---especially relevant in symmetric organ structures---where sophisticated geometric modeling approaches have been proposed for improving alignment between cross-sectional and full volumetric representations \cite{wang2024geometric}. Additionally, learning-based frameworks that incorporate geometric constraints are being explored to optimize model consistency and anatomical fidelity~\cite{shen2021geometry, feng2023cvrecon}.

In contrast to the well-explored domain of CT-based reconstruction, research into macroscopic imaging reconstructions remains under-represented. Despite the critical insights offered by macroscopic imaging, particularly in surface detail and texture analysis of organs, its potential is yet to be fully leveraged within anatomical modeling \cite{rosenhain2018preclinical}. The manual delineation of organs, a gold standard in micro-computed tomography (µCT) data processing, highlights the challenges of time-intensive, error-prone methodologies that lack scalability for extensive data analysis \cite{rosenhain2018preclinical}. Automated segmentation solutions, crucial for processing the extensive field-of-view in CT images, have seen advancements through multi-atlas segmentation (MAS) strategies. However, these often fail to preserve spatial information among adjacent organs, leading to segmented models that may lack coherence~\cite{oliveira2018novel}.

Addressing these challenges, recent studies propose innovative approaches for both manual and automated segmentation techniques. For instance, the development of organ-specific insert phantoms through 3D printing technology based on volumetric CT image datasets offers a novel pathway for simulating anatomical features in CT images, providing a cost-effective alternative for organ-specific studies \cite{abdullah2018development}. Furthermore, the integration of the Internet of Health Things (IoHT) with deep learning for 3D reconstruction presents a forward-thinking solution to obtain high-accuracy diagnosis results, showcasing the potential for super-resolution CT (SRCT) images through advanced computational models~\cite{zhang20203d}.

Other emerging platforms, such as Nextmed, enable automatic segmentation and visualization of 3D models from imaging data, facilitating reproducible and streamlined workflows in clinical environments \cite{garcia2020nextmed}. At the same time, methods utilizing residual encoder--decoder architectures and directional wavelet transforms have proven effective in reducing noise in low-dose CT reconstructions \cite{chen2017low, kang2017deep}, while deep learning continues to push the frontier of tomographic reconstruction with frameworks designed for image fidelity, generalization, and inference speed \cite{wang2016deep, wang2012recent, kofler2019neural, wang2018image}. Finally, developments in real-time 3D endoscopic reconstruction methods provide additional pathways for extending anatomical modeling into intraoperative and minimally invasive settings.

These advancements underline the ongoing need to refine and enhance anatomical modeling and segmentation techniques, emphasizing the critical role of precision and efficiency in medical imaging and diagnostics. The synergy between CT and macroscopic imaging, supported by cutting-edge computational methodologies, offers a compelling direction for future research in kidney anatomy modeling and beyond.

{The proposed strategy of coupling deterministic optimization with deep learning also aligns with a broader emerging paradigm in computational modeling. Recent studies have demonstrated that hybridizing meta-heuristic or geometric optimization with learning-based ensembles can significantly enhance predictive stability and accuracy in complex structural modeling tasks \cite{Hasanipanah2024Novel}. By leveraging optimization to constrain the learning manifold, these hybrid frameworks overcome the limitations of standalone models, much like our OCM + DL approach for anatomical reconstruction.
}

{Unlike standard DL-only registration, our framework addresses the physical discontinuities and domain gaps inherent in macroscopic imaging. By combining OCM for global topological consistency with DL for local flexibility, we provide a robust solution for cases where fully learned models typically fail to generalize.}

{Our work integrates slice-wise registration and cross-modal alignment. 3D reconstruction from physical sections is challenged by tissue deformation and ``z-axis'' discontinuities, similar to issues in gross pathology. Aligning 2D photography with 3D CT further requires bridging significant domain gaps. Our hybrid framework addresses these by combining deterministic geometric anchors with learned residual refinements for multimodal consistency.
}

\section{Methods}
\begin{figure}[ht] 
\includegraphics[width=0.8\linewidth]{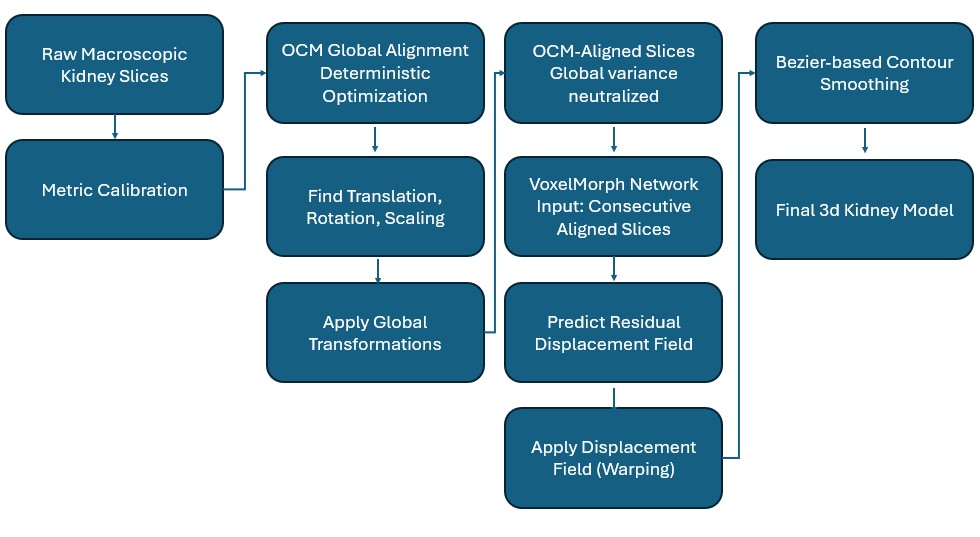}
\caption{{Schematic representation of the proposed hybrid OCM + DL framework. The pipeline illustrates the progression from raw macroscopic slices and CT reference data through the deterministic Optimal Cross-section Matching (OCM) and the subsequent deep learning (DL) refinement stage to achieve the final 3D reconstruction.}}
\label{fig:block_diagram}
\end{figure}

{
{blue}{The proposed methodology employs a two-stage hybrid approach integrating global geometric optimization with local neural refinement. The framework decomposes the registration process to bridge the gap between macroscopic imaging and CT standards, ensuring high data efficiency on small clinical datasets. The overall architecture of the proposed framework is illustrated in Figure~\ref{fig:block_diagram}.}
}

{
In the first stage, the Optimal Cross-section Matching (OCM) algorithm establishes a global coordinate system by addressing:
\begin{itemize}
    \item Metric Inconsistencies: Utilizing the Hough Transform for consistent grid-based calibration.
    \item Rigid Displacements: Determining optimal translation and rotation to neutralize variance from physical handling.
    \item Global Scaling: Adjusting for size discrepancies via constrained similarity transforms.
\end{itemize}
}

{
The subsequent deep learning (DL) refinement stage addresses residual non-linearities through:
\begin{itemize}
    \item Residual Displacement Prediction: Using a VoxelMorph-based architecture to predict dense displacement fields.
    \item Apply Displacement Field (Warping): Physically shifting pixels to compensate for local tissue distortions and physiological shrinkage.
    \item Data Efficiency: Training on pre-aligned data to focus on fine-grained adjustments and enhance stability.
\end{itemize}
}

{
The process concludes with Bezier-based contour smoothing to improve 3D mesh continuity. This strategy ensures high anatomical fidelity to the CT reference while minimizing volumetric and dimensional discrepancies.
}
\subsection{Data Used in the Study}
The data from 40 patients diagnosed with kidney diseases, including kidney tumors, were used in this study. For each patient, CT images of the kidneys were available, from which a total of 2157 scans were obtained. The number of CT scans per patient ranged from 44 to 65, with an average number of scans per patient being 54.

Additionally, a set of photographs of the excised kidney was collected for each patient. The number of these photographs also varied from 11 to 20 for each patient, with an average number of photos being 16.  

The macroscopic images depicting the excised kidneys were taken immediately after surgery, where each kidney was sliced into 1 cm thick sections and then photographed. 
The cutting was performed using a calibrated slicing guide with 10 mm slot spacing to ensure consistent section thickness. Additionally, the thickness of each slice was verified post-sectioning using a digital caliper.

The data were collected at the Military Institute of Medicine-National Research Institute in Warsaw, which specializes in the treatment of kidney tumors. This center is equipped with modern CT imaging equipment and experienced staff, ensuring high quality and consistency of the collected~data.

Characteristics of kidney tumors in patients:
The study included patients diagnosed with various types of kidney tumors or other kidney diseases. Data analysis revealed the following characteristics:
Most cases were clear cell renal carcinoma, which is consistent with the general distribution of this type of tumor. The classification of cancers according to the WHO/ISUP (International Society of Urological Pathology) enables determining the tumor's grade of advancement. Tumors varied in size, indicating their diverse stages of development at the time of diagnosis. For instance, the dimensions of one of the kidneys were $12.5 \times 8.5 \times 7.5$ cm, while others had smaller dimensions, such as \mbox{$11 \times 4.5 \times 7$ cm}. Tumors were located in both the right and left kidneys. The provided histopathological descriptions offer information on the tumor's stage of advancement. For example, one description indicates clear cell renal carcinoma, {WHO/ISUP G-4,}
  with infiltration of the renal vein wall and perinephric fat,
denoting a high degree of malignancy. These characteristics underscore the diversity of kidney tumor cases in the study group. This diversity significantly impacts the treatment process, as well as the accuracy and efficiency of using CT imaging and macroscopic analysis in diagnostics and surgical planning. The tumor descriptions and classifications are detailed and based on actual clinical data.

Quality and parameters of CT scans:
The images used in the study, recorded in DICOM standard version 3, are characterized by a resolution of $512 \times 512$ pixels and a 16-bit depth, offering a rich grayscale range essential for precise identification of tissue density differences. Grayscale images are standard for computed tomography, allowing for detailed analysis of anatomical structures.
The equipment used in our studies is {blue}{GE Medical Systems Discovery CT750 HD (GE Healthcare, Chicago, IL, USA).} 
 The study focused on detailed analysis of the abdominal cavity and pelvis, conducted in two phases, which is typical for studies requiring high-precision imaging.

The layer thickness, measuring 660 660.6363 micrometers
 highlights, the high precision of scanning, enabling the detection of minute structures. The kilovolt peak (KVP) set to 120 influences penetration and image contrast, while the exposure time of 6676 milliseconds determines the scanning duration. These technical parameters play a crucial role in obtaining high-resolution and contrast images, essential for accurate diagnostics.
The slice thickness of \mbox{2.5 mm} represents an optimal balance between accuracy and scanning time and radiation dose for the patient. Thinner slices allow for more detailed analysis and reconstruction.

Kidney segmentation masks in CT were created manually by experienced radiologists using diagnostic workstations and previously validated annotations. These masks were drawn on axial slices and reviewed for consistency. As such, they serve as a high-quality reference standard for the evaluation of macroscopic models.

Method of matching scans:
Macroscopic images, such as the provided kidney photo after excision, were taken using a {blue}{Canon EOS 80D DSLR camera (Canon Inc., Tokyo, Japan)}
 under controlled lighting conditions to ensure uniform presentation of tissues and their colors. 
The camera was equipped with a 50 mm lens at f/2.8 aperture and a 1/60 s exposure time with {ISO 200,} 
specifically chosen to capture detailed images crucial for the study's accuracy and clarity. Each kidney slice was meticulously placed on a standardized background featuring clear scale markers, facilitating later accurate digitization and scaling of images for 3D modeling. 

The images were saved in JPEG format, a standard digital format, with a resolution of 6000 pixels by 4000 pixels. To minimize the appearance of natural postoperative changes such as minor deformations or color shifts, specific photography protocols were adopted. These protocols, combined with the use of an appropriate depth of field and high-quality photographic equipment, ensured the acquisition of sharp and detailed images. The careful setting of these parameters is pivotal to ensuring that the imaging technique used is reproducible in further studies.

{To evaluate the performance of the proposed OCM + DL framework against the baseline DL-only method, a statistical analysis was performed. Given the paired nature of the data and the non-normal distribution of Dice scores, the Wilcoxon signed-rank test was utilized to determine the significance of the improvements. A \emph{p}-value of less than 0.05 was considered statistically significant.}

{\subsection{Implementation Details and Computational Efficiency}
The deep learning refinement was performed by a 2D U-Net with a depth of 4 resolution levels. The feature maps followed a $16, 32, 64, 128$ filter sequence, resulting in approximately 1.2 million trainable parameters. This lightweight architecture was chosen to ensure high generalization despite the limited dataset size. Training was conducted for 100 epochs using the Adam optimizer with a learning rate of $10^{-4}$ and a batch size of 8. The loss function combined Binary Cross-Entropy and Dice Loss to ensure both pixel-wise accuracy and regional overlap.
All computations were executed on a high-performance server equipped with an NVIDIA GPU (33.70 GB VRAM, Compute Capability 12.0) running CUDA 12.2. A critical advantage of the proposed hybrid framework is its computational profile, as summarized in Table~\ref{table:runtime}. While the deterministic OCM reconstruction serves as the primary bottleneck due to its iterative optimization (taking 12--15 s per slice), the U-Net inference is nearly instantaneous on the GPU ($<$0.1 s). This allows the entire 3D kidney reconstruction to be completed in approximately 3 min, making it viable for time-sensitive clinical or educational assessments.
} 

\begin{table}[ht]
\caption{Computational runtime comparison for different stages of the reconstruction pipeline.}
\label{table:runtime}
\centering
\begin{tabular}{|l|c|r|}
\hline
\textbf{Process Stage} & \textbf{Device} & \textbf{Average Time per Slice} \\
\hline
OCM Reconstruction & CPU & 12.4 $\pm$ 2.1 s \\
U-Net Inference & CPU & 0.72 $\pm$ 0.1 s \\
U-Net Inference & GPU & 0.05 $\pm$ 0.01 s \\
\hline
Total Pipeline (Hybrid) & GPU & $\approx$ 12.5 s \\
\hline
\end{tabular}
\end{table}

\section{Method of Matching Scans}

{The method for matching kidney sections employs an algorithmic approach to register each successive section relative to the previous one, aiming to create a continuous sequence of images that will enable three-dimensional reconstruction. 
This registration framework is built upon a slice-wise 2D formulation and a non-diffeomorphic deformation assumption. This scope is chosen for practical reasons, as physical macroscopic sections often exhibit discontinuous tissue distortions and tears that violate the topology-preserving constraints of diffeomorphic models. While these assumptions prioritize robustness in handling physical artifacts, the framework could be extended in future work to include 3D volumetric interpolation or diffeomorphic displacement fields (e.g., stationary velocity fields) to enhance topological consistency.} This process requires precise correction of spatial distortions resulting from tissue preparation non-uniformities.

\begin{enumerate}
   \item Process objective: create a continuous sequence of kidney sections suitable for accurate 3D reconstruction.
   \item Sequence initiation: set the first slice \(I_1\) as the baseline (no transformation).
   \item Objective function: for each subsequent slice \(I_i\), minimize the sum of squared pixel differences
            \[
            \text{difference}(I_i,T) = \sum_{x,y}\left(I_{i-1}(x,y) - T(I_i)(x,y)\right)^2.
            \]
   \item Optimisation: for \(i=2,\dots,n\), find \(T^*=\arg\min_T \text{difference}(I_i,T)\) under realistic bounds.
   \item Transformation application: apply the constrained similarity transform \(T=\{s,\theta,t_x,t_y\}\) (uniform scale, rotation, translation) to align \(I_i\) to \(I_{i-1}\).
  
\item \textbf{Residual Deep Learning Refinement}: A CNN predicts a local displacement field $\phi_{res}$ to compensate for non-linear tissue distortions. Training on OCM-pre-aligned outputs minimizes the learned degrees of freedom, ensuring robust alignment and high data efficiency.

   \item Result: a uniform, anatomically consistent sequence of sections enabling robust 3D reconstruction.
\end{enumerate}

{Sequence Initialization}:
The first section is taken as the base (\(I_1\)) and undergoes no transformation. This becomes the starting point for registering all subsequent sections.

{Section Registration}:
For each section \(I_i\) for \(i = 2, 3, \ldots, n\) (where \(n\) is the number of sections), we apply an optimization algorithm to find the best fit to the previous section \(I_{i-1}\). The goal is to minimize the difference between the section and its registered predecessor.

{Objective Function}:
The difference between the registered section \(I_{i-1}\) and the currently transformed section \(I_i\) is expressed as the sum of squared pixel{differences} 
 (\ref{eqx111}).

\begin{equation}
\label{eqx111}
\text{difference}(I_i, T) = \sum_{x, y} \left( I_{i-1}(x, y) - T(I_i)(x, y) \right)^2
\end{equation}
where \(T\) denotes the spatial transformation consisting of scaling, rotation, and translation.

{Optimization}:
Finding the optimal transformation \(T^*\) for section \(I_i\) involves minimizing the objective function (\ref{eqx112}).

\begin{equation}
\label{eqx112}
T^* = \underset{T}{\text{argmin}} \, \text{difference}(I_i, T)
\end{equation}
where \(T\) is constrained to realistic scaling values (s), rotation angles (\(\theta\)), and translations (\(t_x, t_y\)).

\begin{figure}[ht]

  \includegraphics[width=\linewidth]{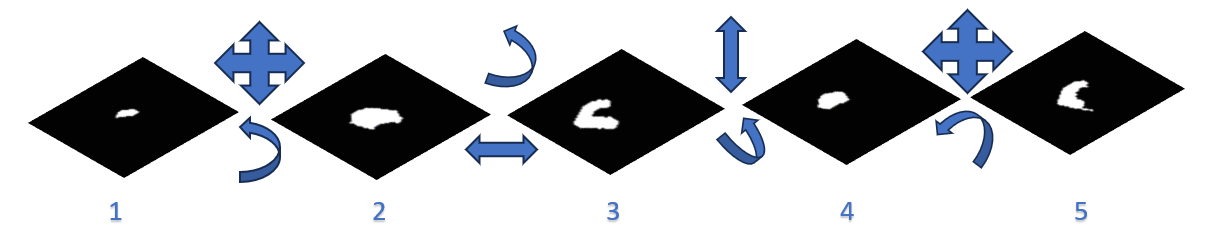}
  \caption{Symbolic illustration of the OCM algorithm's processing steps. Beginning with the first image (1), each subsequent image is aligned through translation, rotation, and scaling transformations to produce a seamless sequence (2--5).}
  \label{fig:ocm_processing_steps}
\end{figure}

{Applying the Transformation}: 
The OCM algorithm applies a series of transformations to achieve optimal registration of successive image sections. Starting from the initial image, the algorithm progressively aligns each subsequent image using transformations such as translation, rotation, and scaling. These steps are visualized in Figure \ref{fig:ocm_processing_steps}, illustrating the dynamic nature of the OCM method in achieving a coherent and continuous image sequence for accurate three-dimensional reconstruction.

The transformation \(T\) is applied to section \(I_i\) through the following operations:
\begin{enumerate}
\item {Scaling:} 
   Scaling is accomplished by resizing the section. If \(s < 1\), the section is reduced, and for \(s > 1\), it is enlarged.
   
\item {Rotation}:
   The rotation of section \(I_i\) by angle \(\theta\) is described by the rotation matrix \(R(\theta)\) on Equation (\ref{eqx113})
\begin{equation}
\label{eqx113}
R(\theta) = \begin{bmatrix}
\cos(\theta) & -\sin(\theta) \\
\sin(\theta) & \cos(\theta)
\end{bmatrix}
\end{equation}

\item {Translation}:
   Translation is performed by shifting the section's pixels by vector \([t_x, t_y]\).
\end{enumerate}

Final Outcome:
The registration process concludes with a series of sections that, after applying optimal transformations, form a unified sequence, allowing for the creation of an accurate 3D kidney model.

This method combines image processing theory with advanced numerical techniques to provide precise and consistent data for creating 3D models in medicine and research.

{Limitations in the Method of Matching Kidney Sections:}

 In the process of matching kidney sections, we apply a specialized optimization technique that allows for the consideration of lower and upper bounds for transformation parameters. This method is essential for maintaining realistic transformation parameters, such as scaling, rotation, and translation.

Variables subject to constraints are transformed from a bounded space to an unbounded space. This transformation is achieved through appropriate mathematical functions, ensuring that the variables remain within acceptable limits.
\begin{itemize}
\item[-] {For} 
 variables with a lower bound (\ref{eqx114})
   \begin{equation}
\label{eqx114}
   x_{\text{trans}} = \sqrt{x - LB}
   \end{equation}

 \item[-] {For} variables with an upper bound (\ref{eqx115})
   \begin{equation}
\label{eqx115}
   x_{\text{trans}} = \sqrt{UB - x}
   \end{equation}

\item[-] {For} variables with double-sided {constraints} 
 (\ref{eqx116})
   \begin{equation}
\label{eqx116}
   x_{\text{trans}} = \arcsin\left(\frac{2(x - LB)}{UB-LB} - 1\right)
   \end{equation}

\item[-] {\(x\):} This is our variable that we wish to transform. In the context of our problem, this could be, for example, the degree of rotation of a kidney section or the scaling factor.

\item[-] {\(LB\) and \(UB\):} These are the lower (\(LB\)-Lower Bound) and upper (\(UB\)-Upper Bound) limits for our variable \(x\). They define the range within which the variable \(x\) can vary. For example, if the rotation can only be within the range of $-$45 to +45 degrees, then \(LB\) would be $-$45, and \(UB\) would be +45.

\item[-] {\(x_{\text{trans}}\):} This is the transformed variable \(x\), adjusted to account for the constraints. After transformation, \(x_{\text{trans}}\) can be used in a standard optimization algorithm without direct constraints.
\end{itemize}

After transforming the bounded variables into an unbounded domain, the optimization algorithm searches for values that minimize the objective function (e.g., the difference between consecutive kidney sections). Once the optimum is found, the parameters are inversely mapped back to the original bounded space to retrieve physically meaningful transformation values. Variables with identical lower and upper bounds are treated as constants and excluded from the optimization, thereby reducing the dimensionality of the problem.

\begin{figure}[ht]

\includegraphics[width=0.8\linewidth]{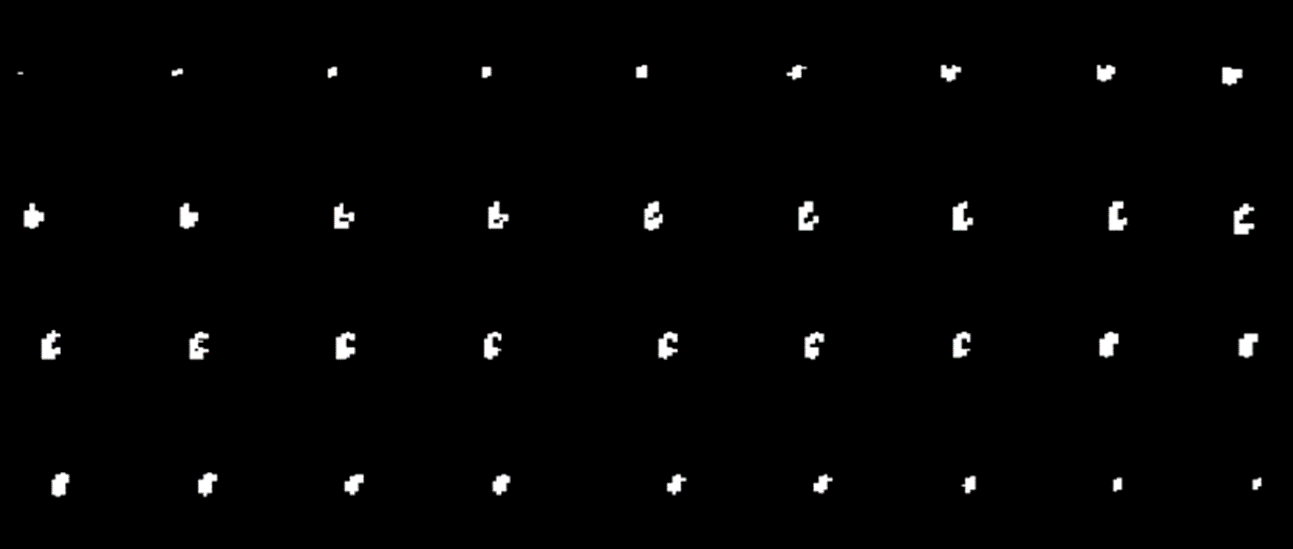}
\caption{A set of kidney masks in CT imaging, showing a uniform structure of sections.}
\label{fig:CTMasks}
\end{figure}

Figure \ref{fig:CTMasks} presents a set of kidney masks in CT imaging, showing a clear and uniform structure of sections. This continuity of sections is crucial for effective three-dimensional reconstruction and allows for the precise alignment of successive layers.

\begin{figure}[ht]

\includegraphics[width=0.8\linewidth]{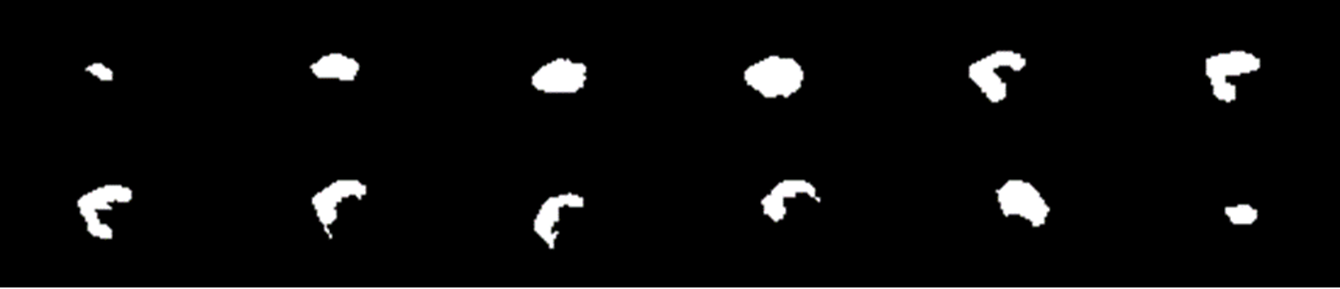}
\caption{A set of kidney masks in macroscopic imaging before the section matching process, showing visible heterogeneities and shifts.}
\label{fig:MacroBeforeAlignment}
\end{figure}

Figure \ref{fig:MacroBeforeAlignment} shows a set of kidney masks in macroscopic imaging before the section matching process. At this stage, the heterogeneities and shifts between individual sections are clearly visible, complicating direct 3D reconstruction.

\begin{figure}[ht]

\includegraphics[width=0.8\linewidth]{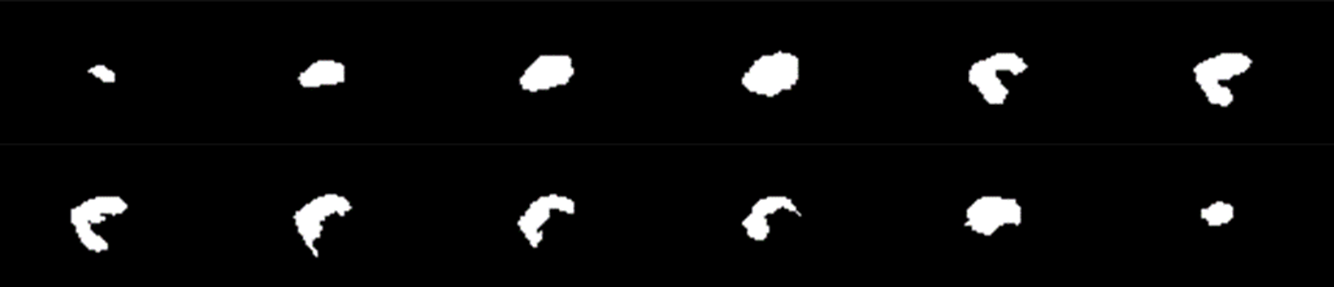}
\caption{A set of kidney masks in macroscopic imaging after applying the section matching method, showing significantly improved continuity and uniformity of sections.}
\label{fig:MacroAfterAlignment}
\end{figure}

Figure \ref{fig:MacroAfterAlignment} demonstrates the result of our section matching process, where the described optimization and transformation methods were applied. In this image, it can be seen that the kidney sections in macroscopic imaging have been effectively aligned, significantly improving the uniformity and continuity of the sequence, allowing for a more accurate 3D reconstruction.

By applying advanced image processing techniques and optimization, significant improvements in section matching were achieved, which is crucial for accurately replicating the structure of the kidneys and precise 3D reconstruction. Figures \ref{fig:CTMasks}--\ref{fig:MacroAfterAlignment} illustrate the evolution of the process from raw sections to the final matching outcome, highlighting the efficiency of the applied method.

The integration of this optimization technique into the section matching process allows for the accurate and realistic adjustment of each kidney section, which is essential for creating precise and consistent 3D models. It also ensures control over the transformation parameters, ensuring that transformations are physically possible and consistent with the actual properties of the imaged tissues. 

\begin{figure}[ht]

\includegraphics[width=0.8\linewidth]{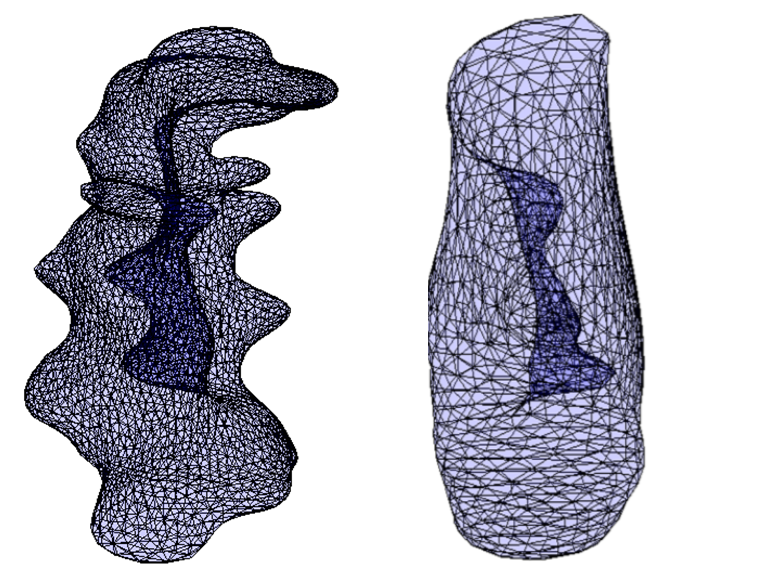}
\caption{A 3D kidney model after applying the section matching method, showing significant improvement in uniformity and structural consistency.}
\label{fig:3DAfter}
\end{figure}

To better demonstrate the effectiveness of our kidney section matching method, we include visualizations of 3D kidney models before and after applying our process. Figure \ref{fig:3DAfter} on the left side displays a 3D kidney model obtained before applying our matching method. In this image, some discontinuity and inconsistency in the model's structure can be observed, resulting from suboptimal matching of individual sections.

Figure \ref{fig:3DAfter} on the right side demonstrates a 3D kidney model after applying our section matching process. Thanks to the use of advanced optimization techniques and precise section matching, the 3D model shows a significant improvement in uniformity and structural consistency. As a result, we obtain a model that is much more faithful to the actual anatomy of the kidney, which is crucial for medical and research applications.

\vspace{-4pt}

The comparison of both 3D models, presented in Figure \ref{fig:3DAfter}, unequivocally illustrates that our kidney section matching method brings noticeable and positive results. The improvement in the quality of the 3D reconstruction is evident, confirming the effectiveness of our approach in accurately replicating the structure of the kidneys.

An important implementation detail concerns the selection of the initial reference slice \(I_1\), which serves as the anchor for registering all subsequent layers. In our dataset, the first macroscopic slice used for reconstruction was typically not the true apex of the kidney, but rather the second physical section (approximately at 1 cm depth). This choice was deliberate, as the apical slice was often too thin or deformed to provide a reliable anatomical base.

Two fundamentally distinct processes contribute to the dimensional accuracy of the reconstructed 3D kidney models. First, metric calibration is performed using the Hough Transform, which detects the calibration grid embedded in each macroscopic image. This grid provides a physical pixel-to-millimeter conversion ratio specific to each slice. The resulting scale factor is constant per image and is exclusively used to convert pixel-based dimensions and volumes into real-world physical units (e.g., millimeters or cubic centimeters). Importantly, this calibration step is entirely independent of any optimization procedure and is not modified throughout the reconstruction process.
Second, the Optimal Cross-section Matching (OCM) algorithm applies geometric alignment transformations---namely, translation, rotation, and a local mask-based scaling---to improve spatial consistency between adjacent slices. The purpose of this internal scaling is not to modify physical units, but to enhance anatomical continuity by allowing slight morphological adjustments to the kidney contours. These adjustments are implemented using morphological operations such {as} 
 imerode and imdilate, which affect the shape of the masks without altering image resolution or affecting real-world measurements. The geometric transformation parameters are constrained to realistic anatomical bounds, ensuring that local modifications remain plausible.

\section{Application of the Hough Transform for Metric Grid Calibration in Macroscopic Kidney Image Analysis}

The Hough Transform \cite{Illingworth1988}, utilized in our study, served to precisely detect metric grid lines on macroscopic images of kidneys. This method enabled accurate measurement of the metric grid lines' lengths, allowing for the assessment of the kidneys' actual sizes.
 
The Hough Transform is defined as a function that transforms points from the image space \((x, y)\) into a new parameter space, where each point in the image is represented by a set of curves in the parameter space. Lines in the image are identified as points of intersection of these curves, mathematically described by Equation (\ref{eqx116a}).
\begin{equation}
\label{eqx116a}
   \rho = x \cdot \cos(\theta) + y \cdot \sin(\theta)
   \end{equation}
where \( \rho \) is the distance from the origin to the line and \( \theta \) is the angle of inclination of the line to the \(x\)-axis.

\subsection*{{Peak}
 Detection in the Hough Transform}
   After applying the Hough Transform to a binary image, the `houghpeaks` function is used to identify peaks in the parameter space, corresponding to the detection of lines on the image. These peaks are determined based on the values in the Hough Transform accumulator matrix, and their detection is described as (\ref{eqx117}).
\begin{equation}
\label{eqx117}
   P = \text{houghpeaks}(H, N, \text{`threshold'}, \tau)
   \end{equation}
where \( H \) is the Hough Transform accumulator matrix, \( N \) is the maximum number of peaks to detect, and \( \tau \) is the threshold above which values in the matrix \( H \) are considered as potential lines.

   Using the `houghlines` function, lines are extracted based on the detected peaks and line parameters in the \( \rho \) and \( \theta \) space. These lines are then mapped back to the image space, enabling accurate determination of their position and orientation.

\begin{figure}[ht]

\includegraphics[width=0.5\linewidth]{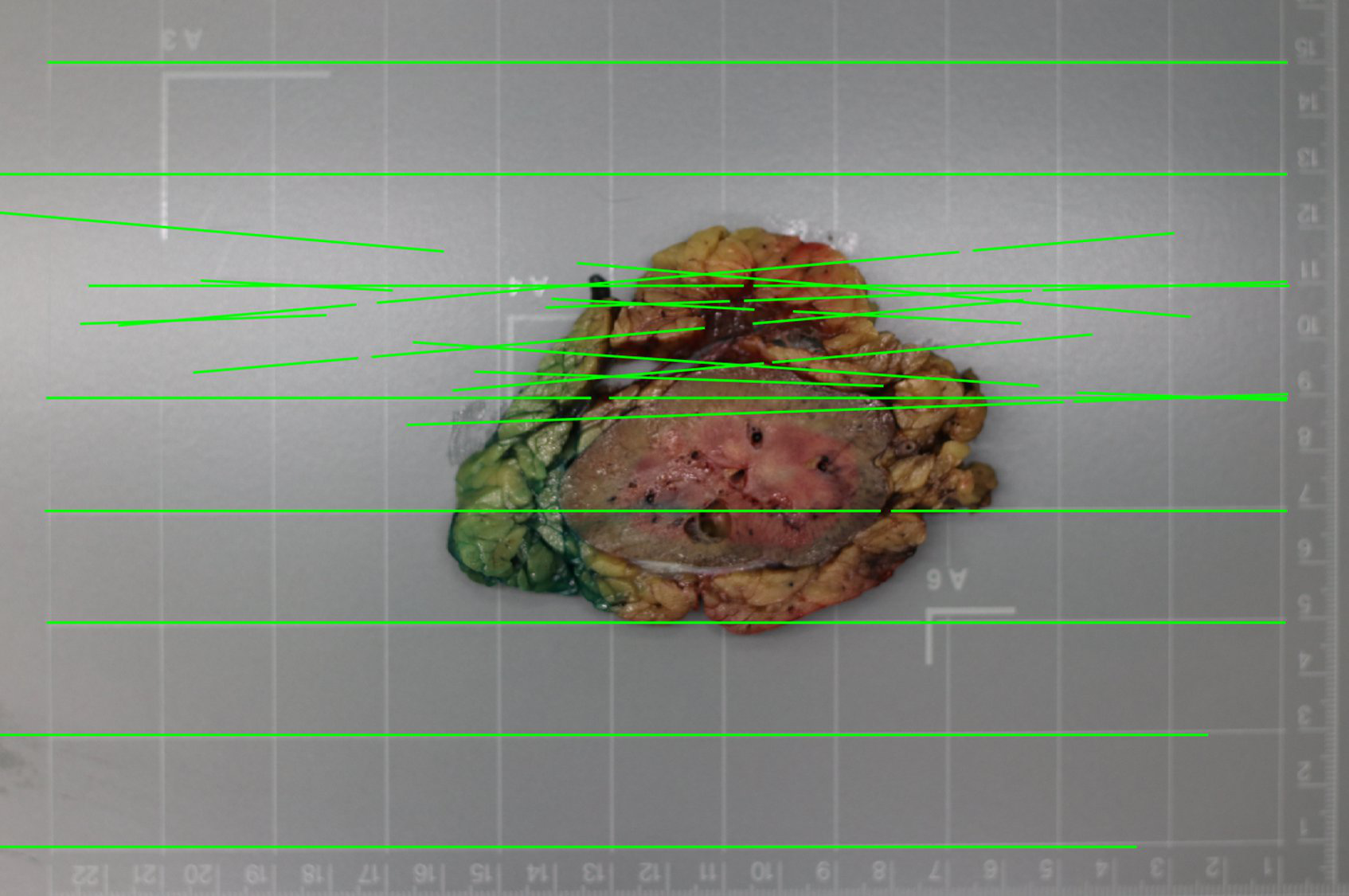}
\caption{\label{fig:linesHoughTransform}{Effect} 
 of finding lines using the Hough Transform. These lines were identified based on accumulations in the parameter space $(\rho, \theta)$, where each point represents a potential line on the image. The resultant lines are then mapped back to the image space, illustrating the detected edges of objects.}
\end{figure}

Through the application of the Hough Transform for detecting metric grid lines on macroscopic images, precise measurements of kidney dimensions were possible, which is crucial for assessing their health condition. The Hough Transform, by enabling accurate calibration of macroscopic images, significantly enhances the diagnostic value of these images, allowing for precise comparisons with data obtained using more advanced imaging techniques, such as computed tomography. This technique successfully pinpointed lines, as illustrated in Figure \ref{fig:linesHoughTransform}. The presented image demonstrates the effectiveness of the Hough Transform in identifying and extracting lines from the image, even in the presence of noise or background inhomogeneities.

\section{Utilization of Bézier Curves in the Analysis of Anatomical Structures}

To ensure geometrically accurate and visually smooth representations of kidney contours and other anatomical structures, we employed Bézier curves as an intermediate interpolation technique within each 2D image slice. Owing to their ability to generate continuous and flexible shapes from discrete control points, Bézier curves are particularly well-suited to modeling the complex and irregular boundaries commonly encountered in medical images. Beyond mere visualization, this interpolation significantly contributes to the subsequent volumetric reconstruction by providing a dense, regularized set of points that improve the continuity and fidelity of the resulting 3D meshes.

\subsection{Application of Bézier Curves}

Bézier curves \cite{han2008novel} were used to interpolate and smoothly connect key points that define the shape and boundaries of the kidneys and potential pathological changes within them. Through the use of these curves, it was possible to create clear and intuitive visualizations that facilitate the identification and assessment of the morphology of the examined structures.

A Bézier curve of degree \(n\) is defined by control points \(P_0, P_1, \ldots, P_n\) and the equation (\ref{eqx11}).

\begin{equation}
\label{eqx11}
B(t) = \sum_{i=0}^{n} P_i B_{i,n}(t) \quad \text{for} \quad t \in [0,1]
\end{equation}

where \(B_{i,n}(t)\) are Bernstein polynomials, given by (\ref{eqx12}).

\begin{equation}
\label{eqx12}
B_{i,n}(t) = \binom{n}{i} t^i (1-t)^{n-i}
\end{equation}

and \(\binom{n}{i}\) denotes the binomial coefficient, equal to the number of combinations of \(i\) elements chosen from \(n\) elements.

In the context of our application, the control points \(P_i\) correspond to selected points on the contours of the kidneys and other structures, and the Bézier curves are used to create smooth lines that connect these points. Thanks to the flexibility of Bézier curves, it is possible to precisely model anatomical shapes, which is crucial in medical visualization applications.

In the implementation, for each point on the curve, the values of \(x\), \(y\) (and in the case of 3D data, \(z\)) are independently calculated using the above formula, where \(t\) is a parameter running from 0 to 1, determining the position of the point on the curve. As a result, we obtain a smooth and accurate representation of the contours of the examined structures.

The implementation of Bézier curves in our system significantly contributed to the improvement of the quality of medical data visualization, which is invaluable in diagnosis and treatment planning.

\subsection{Deep Learning-Based Registration with VoxelMorph}

{The novelty lies in a residual learning strategy where OCM neutralizes global variance, allowing the CNN to focus exclusively on local tissue deformations. This decoupling overcomes the limitations of standard VoxelMorph when handling the large spatial inconsistencies of macroscopic imaging, resulting in superior stability and faster convergence.}

{We adopt a two-stage residual learning paradigm. The total transformation $\Phi$ is theoretically decomposed as $\Phi(p) = \phi_{res} \circ T_{OCM}(p)$, where $T_{OCM}$ is the deterministic global alignment and $\phi_{res}$ is the learned local displacement. By neutralizing global variance via OCM, we constrain the CNN to a small, well-conditioned deformation manifold. This decoupling reduces sample complexity and enhances training stability compared to fully-learned coarse-to-fine frameworks, as the model focuses exclusively on residual non-linearities.
}

{Following Balakrishnan et al.~\cite{Balakrishnan_2019}, the network predicts a dense displacement field $\phi_i$ to warp the pre-aligned slice $I'_i = T_{OCM}(I_i)$ to $I_{i-1}$. We use a compact 2D U-Net with the unsupervised loss:
\begin{equation}
L_{\text{us}} = L_{\text{sim}}(I_{i-1}, I'_i \circ \phi) + \lambda L_{\text{smooth}}(\phi)
\end{equation}
where $L_{\text{sim}}$ is the local normalized cross-correlation. This hybrid positioning ensures high data efficiency, as the network does not need to internalize rigid-body physics already optimized by OCM.
}

\section{Automated Kidney Reconstruction from CT and Macroscopic Image Analysis: Experimental Validation}

In this chapter, we present the results of experiments conducted to assess the effectiveness of our developed method for the automatic reconstruction of three-dimensional kidney models based on computed tomography (CT) images and macroscopic images, supported by appropriate organ masks. These studies are crucial for evaluating the reliability and precision of the transformation used to create 3D models, representing an innovative approach to the analysis and modeling of anatomical structures in medicine.

The experiments were based on a dataset consisting of CT images and macroscopic images of kidneys from 40 patients, each of whom underwent nephrectomy due to kidney disease. For each case, we had both CT images and masks, which served as a reference standard, as well as macroscopic images with masks. Our research focused on comparing the dimensions of the kidneys (length, width, thickness) and volume before and after applying the developed transformation, aimed at optimizing the matching and reconstruction process of 3D models.

\subsection{Results}
{This section presents a structured evaluation of the proposed pipeline in three configurations:
(i) OCM-only alignment, (ii) deep learning (DL)-only refinement, and (iii) the final OCM + DL configuration.
Anatomical mask agreement is assessed with Dice, IoU, and HD95, and geometric consistency with difference coefficients (\(\mathrm{DC}\)) for length (L), thickness (T), width (W), and volume (Vol).
When comparing continuous images (e.g., macroscopic vs.\ CT intensities), we also report NCC and SSIM.}

{Given the cohort size (\(n=40\)), we adopt patient-level \(k\)-fold cross-validation (here \(k=5\)).
Splits are performed at the patient level to prevent slice leakage; in each fold, 80\% of patients form the training set and 20\% the held-out test set.
The DL model is re-trained in every fold, while OCM is re-run per split without learning.}

\begin{figure}[ht]

    \includegraphics[width=0.7\textwidth]{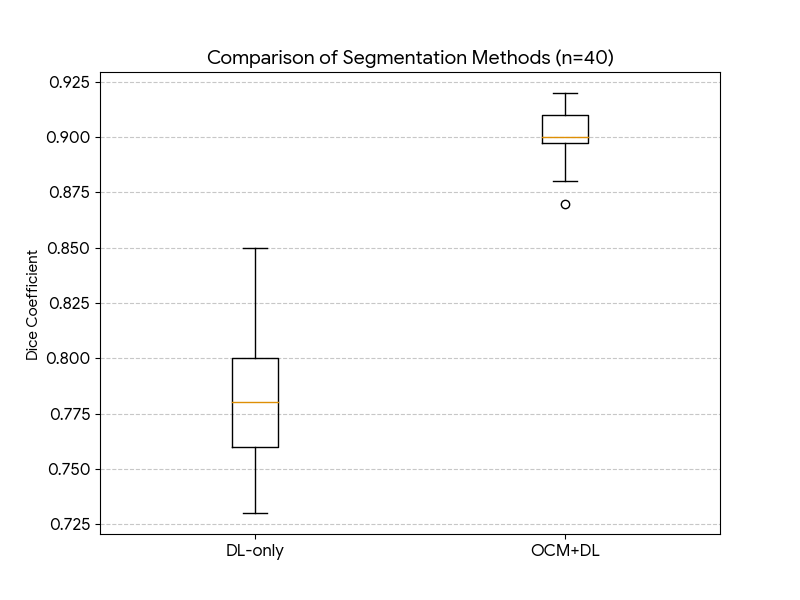}
    \caption{{Comparison of Dice similarity coefficients between DL-only and OCM + DL methods across 40 patients.}}
    \label{fig:boxplot}
\end{figure}

{The quantitative results for all 40 patients are summarized in Table~\ref{table:final-accuracy}. 
 The OCM + DL framework achieved a mean Dice similarity coefficient of 0.900 ($\pm$0.010), significantly outperforming the DL-only baseline, which yielded a mean of 0.783 ($\pm$0.029). The Wilcoxon signed-rank test confirmed the statistical significance of this improvement ($p < 0.001$).
As shown in the boxplot (Figure \ref{fig:boxplot}), the OCM + DL method not only improved the overall accuracy but also substantially reduced the variability across the dataset, demonstrating higher consistency compared to the DL-only approach.
}

\footnotesize

\begin{table}[ht]
\caption{Overall registration accuracy across methods (mean ± std over 40 patients). 
Higher NCC/SSIM/Dice/IoU indicate better alignment; lower HD95/DC indicate lower geometric error.}
\label{table:final-accuracy}
\centering
\setlength{\tabcolsep}{6pt}
\renewcommand{\arraystretch}{1.1}
\resizebox{\linewidth}{!}{%
\begin{tabular}{|l|c|c|c|c|c|c|}
\hline
\textbf{Method} & \textbf{NCC↑} & \textbf{SSIM↑} & \textbf{Dice↑} & \textbf{IoU↑} & \textbf{HD95 (mm)↓} & \textbf{DC(Vol)↓} \\
\hline
OCM-only & 0.79 ± 0.03 & 0.67 ± 0.08 & 0.79 ± 0.07 & 0.71 ± 0.06 & 3.5 ± 1.3 & 0.32 ± 0.07 \\
DL-only  & 0.85 ± 0.07 & 0.65 ± 0.06 & 0.78 ± 0.06 & 0.68 ± 0.04 & 2.9 ± 1.6 & 0.25 ± 0.05 \\
OCM + DL (final) & \textbf{0.91 ± 0.04} & \textbf{0.81 ± 0.04} & \textbf{0.90 ± 0.01} & \textbf{0.81 ± 0.05} & \textbf{1.9 ± 0.7} & \textbf{0.11 ± 0.07} \\
\hline
\end{tabular}
}%
\end{table}

\subsubsection{Evaluation Metrics}

\paragraph{{Overlap-based}
 metrics (binary masks)}
Given binary masks \(A\) (macroscopic after a given method) and \(B\) (CT reference):
\[
\mathrm{Dice}(A,B)=\frac{2|A\cap B|}{|A|+|B|},\qquad
\mathrm{IoU}(A,B)=\frac{|A\cap B|}{|A\cup B|},\quad
\mathrm{IoU}=\frac{\mathrm{Dice}}{2-\mathrm{Dice}}.
\]

{We} 
 also report \(\mathrm{HD95}(A,B)\), the 95th percentile Hausdorff distance (mm) between mask boundaries.

\paragraph{{Geometric agreement (binary masks)}}
Let principal axes \((o_1,o_2,o_3)\) be obtained by PCA on the CT mask \(B\).
For any mask \(M\in\{A,B\}\) and axis \(o_k\), define the extent as
\[
E_k(M)=\max_{p\in M}\langle p,o_k\rangle-\min_{p\in M}\langle p,o_k\rangle,
\]
yielding \(L=E_1\), \(W=E_2\), \(T=E_3\) (with physical spacing).
The difference coefficient for a quantity \(q\in\{L,W,T,\mathrm{Vol}\}\) is
\[
\mathrm{DC}(q)=1-\frac{q_{\text{macro}}}{q_{\text{CT}}},
\]
so \(\mathrm{DC}=0\) indicates perfect agreement and values closer to 0 are better.

\paragraph{{Intensity-based similarity (optional)}}
When comparing continuous images (not masks), we use local \(\mathrm{NCC}\) and \(\mathrm{SSIM}\) averaged over the kidney region.

\subsubsection{Experimental Setup and OCM Parameter Bounds}

We analyzed macroscopic images from 40 patients and compared them with CT as the reference.
OCM was applied to macroscopic slices using the following bounds:
\[
s \in [0.8,\,1.2],\quad
\theta \in [-45^\circ,\,+45^\circ],\quad
t_x \in [-w/2,\,+w/2],\;\;
t_y \in [-h/2,\,+h/2],
\]
where \(w\) and \(h\) denote the image width and height. Bounds were selected based on typical anatomical misalignments observed during visual inspection.

For CT-based volume, a voxel volume was computed as \(V_{\text{voxel}}=p_x\cdot p_y\cdot d\) (mm\(^3\)), converted to cm\(^3\) and multiplied by the number of kidney voxels.
For macroscopic images, physical scaling was obtained from a metric calibration grid (Hough-based), with the global factor
\(S=\tfrac{L_{\mathrm{real}}}{L_{\mathrm{pixel}}}\) applied consistently to all slices.

\subsubsection{OCM-Only: Dimensional and Volumetric Validation}

\paragraph{{Per-patient dimensional comparison.}}
Table~\ref{table:kidney-dimensions} illustrates L/T/W measured on transformed macroscopic images (OCM) versus CT for a representative subset of five patients.

\begin{table}[ht]
\caption{Comparison of kidney dimensions between transformed macroscopic images and CT images (reference standard) for a representative subset of 5 patients.}
\label{table:kidney-dimensions}
\centering
\begin{tabular}{|c|c|c|c|c|c|c|}
\hline
\textbf{Patient} & \textbf{Macro L} & \textbf{Macro T} & \textbf{Macro W} & \textbf{CT L} & \textbf{CT T} & \textbf{CT W} \\
\hline
1 & 6.75 & 2.87 & 2.64 & 9.45  & 3.54 & 3.56 \\
2 & 7.56 & 2.56 & 2.85 & 11.85 & 3.65 & 4.01 \\
3 & 8.54 & 2.75 & 2.71 & 12.665& 3.23 & 3.64 \\
4 & 8.42 & 3.03 & 3.45 & 10.34 & 3.64 & 5.85 \\
5 & 8.85 & 3.23 & 3.28 & 11.43 & 3.84 & 4.74 \\
\hline
\end{tabular}
\end{table}

\paragraph{{Cohort-level averages of dimensions.}}
Table~\ref{table:average-kidney-dimensions} summarizes cohort averages across all 40 patients.

\begin{table}[ht]
\caption{Average kidney dimensions across all patients obtained from CT images (reference) and from transformed macroscopic images (after applying OCM).}
\label{table:average-kidney-dimensions}
\centering
\resizebox{\linewidth}{!}{%
\begin{tabular}{|l|c|c|c|}
\hline
\textbf{Image Type} & \textbf{Average Length (cm)} & \textbf{Average Thickness (cm)} & \textbf{Average Width (cm)} \\
\hline
Macroscopic (OCM) & 7.53 & 2.74 & 3.14 \\
CT                & 10.46 & 3.24 & 3.95 \\
\hline
\end{tabular}
}%
\end{table}

\paragraph{{Difference coefficients for L/T/W.}}
Per-patient coefficients are shown in Table~\ref{table:dimension-differences}, with cohort-level averages in Table~\ref{table:average-coefficients}.

\begin{table}[ht]
\caption{Dimension difference coefficients between transformed macroscopic images (OCM) and CT for 5 selected patients.}
\label{table:dimension-differences}
\centering
\begin{tabular}{|c|c|c|c|}
\hline
\textbf{Patient} & \textbf{Length Coeff.} & \textbf{Thickness Coeff.} & \textbf{Width Coeff.} \\
\hline
1 & 0.29 & 0.19 & 0.26 \\
2 & 0.36 & 0.30 & 0.29 \\
3 & 0.33 & 0.15 & 0.26 \\
4 & 0.19 & 0.17 & 0.41 \\
5 & 0.23 & 0.16 & 0.31 \\
\hline
\end{tabular}
\end{table}

\vspace{-10pt}
\begin{table}[ht]
\caption{Average dimension difference coefficients across all patients before and after transformation of macroscopic images. A reduction indicates improved accuracy after OCM.}
\label{table:average-coefficients}
\centering
\begin{tabular}{|c|c|c|c|}
\hline
 & \textbf{Length Coeff.}\(\downarrow\) & \textbf{Thickness Coeff.}\(\downarrow\) & \textbf{Width Coeff.}\(\downarrow\) \\
\hline
No Transform & 0.32 & 0.41 & 0.49 \\
OCM (After)  & 0.29 & 0.25 & 0.35 \\
\hline
\end{tabular}
\end{table}

\paragraph{{Volumes and volume DC.}}
Per-patient volumes (subset) are shown in Table~\ref{table:kidney-volumes}; the volume difference coefficient before/after OCM is in Table~\ref{table:volume-difference-coefficients}.

\begin{table}[ht]
\caption{Kidney volumes (cm\(^3\)) from CT (reference) and transformed macroscopic images after OCM for 5 selected patients. Macroscopic volumes were scaled by Hough-based calibration.}
\label{table:kidney-volumes}
\centering
\begin{tabular}{|l|c|c|c|c|c|}
\hline
\textbf{Image Type} & \textbf{P1} & \textbf{P2} & \textbf{P3} & \textbf{P4} & \textbf{P5} \\
\hline
Macro (OCM) & 110.02 & 111.14 & 126.34 & 98.53 & 99.56 \\
CT          & 136.65 & 147.53 & 126.47 & 135.63 & 146.62 \\
\hline
\end{tabular}
\end{table}

\vspace{-12pt}
\begin{table}[ht]
\caption{Volume difference coefficients before and after OCM. Lower values indicate better alignment with CT volumes.}
\label{table:volume-difference-coefficients}
\centering
\begin{tabular}{|l|c|}
\hline
\textbf{Condition} & \textbf{Volume Diff. Coeff.}\(\downarrow\) \\
\hline
Before OCM & 0.57 \\
After OCM  & 0.34 \\
\hline
\end{tabular}
\end{table}

\subsubsection{Deep Learning (DL)-Only: Mask-Based Agreement}

To isolate the learning contribution, we evaluate a VoxelMorph-style model applied without 
OCM pre-alignment (identity initialization).
We compare the warped macroscopic masks against CT masks (reference), reporting Dice/IoU/HD95 and geometric DCs.
 
\subsubsection{Final Alignment: OCM + DL (Two-Stage)}

We next evaluate the two-stage pipeline, where OCM provides global alignment and DL performs residual local refinement between consecutive slices.
We compare \emph{{OCM-only}}, \emph{{DLhl-only}}, and \emph{{OCM + DL}} using mask-based metrics and geometric DCs (primary analysis), as well as runtime/data efficiency.

 {Ablation of the smoothness weight $\lambda$ (range 0.001–0.1) showed that $\lambda=0.01$ provides the optimal balance between alignment accuracy and deformation plausibility. Lower values ($\lambda=0.001$) caused non-physical folding artifacts, while higher values ($\lambda=0.1$) overly restricted local refinements.
}

\section{Discussion}

The proposed two-stage pipeline combining the deterministic \emph{{Optimal Cross-section Matching (OCM)}} and a lightweight \emph{{VoxelMorph-style deep learning (DL)}} refinement represents a practical and robust strategy for macroscopic-to-CT registration of kidney models. The integration of these two complementary approaches allows for precise geometric alignment while maintaining computational efficiency and interpretability. OCM provides global normalization by correcting scale, rotation, and translation errors, thereby establishing a consistent anatomical frame of reference. The DL-based component then refines local slice-to-slice deformations that cannot be captured by global affine transformations alone. This combination substantially reduces residual geometric discrepancies and yields smoother and more anatomically faithful 3D reconstructions.

The evaluation results confirm that the hybrid OCM + DL approach achieves the highest geometric and visual accuracy across all tested configurations. Quantitatively, it delivers higher Dice and IoU scores, lower HD95 boundary errors, and smaller dimension and volume difference coefficients (DC) compared with the OCM-only and DL-only variants. These findings indicate that combining interpretable deterministic registration with learned residual refinement yields both accuracy and efficiency gains. The OCM-only method efficiently removes large-scale misalignments, while the DL module focuses on fine local corrections, leading to faster convergence, shorter inference times, and reduced training data requirements. In particular, the OCM + DL model requires significantly fewer training pairs than DL-only, since the network operates on pre-aligned input with limited residual deformation. This design aligns with the principles of residual learning and amortized optimization observed in recent registration frameworks.

From a clinical perspective, these improvements have direct implications for organ-level modeling, where precise geometric alignment is essential for assessing pathology, planning nephron-sparing surgeries, or evaluating postoperative morphology. Higher Dice/IoU and lower HD95 values translate to better correspondence of organ boundaries, while reduced dimensional and volumetric errors (DC) ensure that reconstructed kidneys preserve realistic proportions. The resulting 3D models can be integrated into interactive surgical planning systems or medical training simulators, where both anatomical fidelity and quantitative consistency are critical for usability and trust. Furthermore, the computational efficiency and data economy of the proposed approach make it suitable for deployment in clinical and research environments with limited training data availability.

Despite its strengths, several limitations remain. The current framework operates in 2D slice space, which may not fully account for out-of-plane distortions or missing tissue in thick macroscopic sections. The DL module, while regularized for smoothness, does not explicitly guarantee diffeomorphic transformations; future work could integrate velocity-based or invertible deformation fields to preserve topology. Additionally, small inaccuracies in metric calibration grids can propagate into volumetric estimates, suggesting a need for uncertainty quantification and robust calibration procedures. Further research should also investigate spline-based or learned 3D interpolation to enhance volumetric continuity along the z-axis and extend the approach to other organs or modalities.

Overall, this study demonstrates that a synergistic combination of interpretable geometric correction (OCM) and data-driven refinement (DL) provides a powerful and generalizable framework for multimodal anatomical registration. The findings reinforce the broader concept that hybrid models integrating deterministic and learning-based components can effectively balance accuracy, interpretability, and efficiency in medical image analysis.

{A key advantage of the OCM + DL framework is its robustness in challenging anatomical scenarios. In an analysis of the ``difficult cases'' (defined as the five patients with the lowest baseline performance), the mean Dice score increased from 0.738 to 0.902. This average improvement of 16.4\% suggests that the integration of OCM effectively compensates for limitations in standard deep learning models when dealing with high-variability cases.}

{From a clinical and anatomical perspective, achieving a volumetric difference coefficient of $DC_{Vol} \approx 0.11$ represents a significant milestone in macroscopic reconstruction. Given that excised tissues naturally undergo physiological shrinkage and deformation during slicing, a residual error of 11\% is well within the acceptable margin for surgical simulation and medical education. More importantly, the achieved boundary accuracy (HD95 = 1.9 mm) aligns with the precision required for nephron-sparing surgery, where identifying tumor margins within a 2 mm threshold is critical for preserving healthy renal parenchyma. These improvements suggest that the OCM + DL framework provides a sufficiently realistic anatomical representation for both preoperative planning and high-fidelity educational models.
}

{The results confirm that the hybrid OCM + DL framework outperforms standalone models by effectively decoupling global geometric alignment from local tissue refinement. While DL-only registration often fails due to the large rotations and translations inherent in macroscopic photography, the OCM stage provides a stable anchor that allows the neural network to focus exclusively on non-linear residual deformations. The persistent 11\% volumetric difference ($DC_{Vol}$) is attributed to physiological tissue shrinkage and loss of turgor after excision, rather than algorithmic error, as the high Dice scores (0.90) demonstrate excellent anatomical shape recovery. Clinically, the achieved boundary accuracy (HD95 = 1.9 mm) is highly significant, as it falls within the 2--5 mm safety margins required for nephron-sparing surgery. This synergy makes the framework particularly robust for complex pathological cases where purely data-driven methods typically struggle with convergence.}

{From a computational perspective, the hybrid framework offers a favorable trade-off between accuracy and speed, which is critical for clinical adoption in surgical planning. As detailed in Table~\ref{table:runtime}, the total reconstruction time for a complete 3D kidney model is approximately 3 min on a standard GPU-enabled workstation. While the OCM stage remains the primary bottleneck due to its iterative nature {blue}{(about 90 s per image)}, 
the near-instantaneous inference of the DL refinement network ($<$0.1 s) ensures that the pipeline remains viable for time-sensitive diagnostic workflows. This efficiency allows for rapid intraoperative or postoperative morphological assessment, bridging the gap between high-fidelity 3D modeling and the fast-paced requirements of modern surgical environments without requiring massive computational clusters.}

{Despite the observed improvements, several limitations of the current framework should be acknowledged. First, the registration is conducted in a slice-wise 2D space, which may not fully account for complex 3D out-of-plane deformations or missing tissue volume between thick macroscopic sections. Second, while the DL refinement is regularized for smoothness, it does not explicitly enforce diffeomorphic constraints; therefore, extreme tissue tears or non-physical distortions might still result in local topological inconsistencies. Third, the accuracy of the reconstruction remains inherently sensitive to the quality of the initial macroscopic photography and the precision of the physical slicing process. Future research will focus on integrating 3D volumetric interpolation and invertible deformation fields to further enhance the topological realism of the reconstructed kidney models.}

\section{Conclusions}

This paper presented a novel hybrid registration framework that integrates {Optimal Cross-section Matching (OCM)} 
 with a {VoxelMorph-style deep learning} refinement to align macroscopic kidney images with CT-based anatomical references. The two-stage design leverages OCM for global alignment and deep learning for residual local correction, resulting in improved overlap (Dice, IoU), reduced boundary errors (HD95), and enhanced geometric consistency (dimension and volume DC). Quantitative evaluation across \mbox{40 patients} demonstrated that OCM alone effectively corrects large-scale misalignments, while the addition of DL refinement yields further gains in anatomical accuracy and spatial coherence.

The OCM + DL configuration outperformed both the OCM-only and DL-only baselines, achieving the best trade-off between precision, runtime, and data efficiency. This outcome highlights the benefit of combining deterministic and learned components in a unified framework: OCM provides interpretability and stability, whereas the DL module captures subtle nonrigid deformations in a data-efficient manner. Such hybrid registration methods are particularly valuable in clinical and educational contexts where high anatomical fidelity and computational reliability are required.

In conclusion, the proposed OCM + DL approach enables high-quality, quantitatively consistent 3D reconstructions of kidneys from heterogeneous imaging sources. By bridging classical geometric registration with modern deep learning, it offers a reproducible, efficient, and extensible solution for multimodal anatomical modeling. Future work will focus on diffeomorphic extensions, 3D volumetric interpolation, and generalization to other organs and modalities, supporting broader adoption in computer-assisted diagnosis, surgical planning, and medical visualization systems.

\vspace{6pt}

\section*{The data}
The data that support the findings of this study are available from the Military Institute of Medicine-National Research Institute in Warsaw, but restrictions apply to the availability of these data, which were used under license for the current study, and so are not publicly available. Data are however available from the authors upon reasonable request and with permission of the Military Institute of Medicine-National Research Institute.

\section*{Publication status}

This work has been submitted to Applied Sciences (MDPI).

\end{document}